\documentclass[10pt,journal,compsoc]{IEEEtran}
\ifCLASSOPTIONcompsoc
  \usepackage[nocompress]{cite}
\else
  \usepackage{cite}
\fi
\ifCLASSINFOpdf
  \usepackage[pdftex]{graphicx}
\else
\fi
\usepackage{dcolumn}% Align table columns on decimal point
\usepackage{bm}% bold math
\usepackage{algorithm}
\usepackage[noend]{algpseudocode}
\usepackage{color}
\usepackage{lipsum}
\usepackage{mathtools}
\usepackage{cuted}

\usepackage{amsmath}
\usepackage{amsthm}
\newtheorem{definition}{Definition}
\newtheorem{theorem}{Theorem}
\usepackage{acronym}
\newcounter{tempEquationCounter} 
\newcounter{thisEquationNumber}
\newenvironment{floatEq}
{\setcounter{thisEquationNumber}{\value{equation}}\addtocounter{equation}{1}% record equation as happened and remember number
\begin{figure*}[!t]% float following equation across columns
\normalsize\setcounter{tempEquationCounter}{\value{equation}}% record current equation number in floated location
\setcounter{equation}{\value{thisEquationNumber}}% use previous equation number
}
{\setcounter{equation}{\value{tempEquationCounter}}% set back to equation number in floated location
\hrulefill\vspace*{4pt}% add a horizontal rule separator
\end{figure*}% end float environment

}

\newcommand{\NEW}[1]{\textcolor{black}{#1}}

\usepackage{hyperref}

\usepackage{atbegshi,picture}
\AtBeginShipout{\AtBeginShipoutUpperLeft{%
  \put(\dimexpr\paperwidth-0cm\relax,-0.3cm){\makebox[0pt][r]{\makebox{\small{This is the author's version of an article that has been published in this journal. Changes were made to this version by the publisher prior to publication.}}}}%
}}
\AtBeginShipout{\AtBeginShipoutUpperLeft{%
  \put(\dimexpr\paperwidth-5cm\relax,-0.6cm){\makebox[0pt][r]{\makebox{\small{The final version of record is available at \url{http://dx.doi.org/10.1109/TNSE.2019.2949036}}}}}%
}}

\usepackage{xcolor}
\usepackage[normalem]{ulem}
\usepackage{enumitem}
\newcommand{\ashkan}[1]{{\color{black} #1}}
\begin{document}
\title{Community Detection and Improved Detectability in Multiplex Networks}

\author{Yuming Huang$^1$}

% \altaffiliation[Also at ]{Physics Department, North Carolina State University.}%Lines break automatically or can be forced with \\
\author{Ashkan Panahi$^2$}%
\author{Hamid Krim$^2$}
\author{Liyi Dai$^3$}

\author{Yuming Huang,
        Ashkan Panahi,~\IEEEmembership{Member,~IEEE,}
        Hamid Krim,~\IEEEmembership{Fellow,~IEEE,}
        and~Liyi Dai,~\IEEEmembership{Fellow,~IEEE}% <-this % stops a space
% \IEEEcompsocitemizethanks{\IEEEcompsocthanksitem Hamid Krim was with the Department
% of Electrical and Computer Engineering, North Carolina State University, Raleigh,
% NC.\protect\\
% % note need leading \protect in front of \\ to get a newline within \thanks as
% % \\ is fragile and will error, could use \hfil\break instead.
% E-mail: yhuang26@ncsu.edu
% \IEEEcompsocthanksitem J. Doe and J. Doe are with Anonymous University.}% <-this % stops a space
% \thanks{Manuscript received April 19, 2005; revised August 26, 2015.}
}

\markboth{IEEE TRANSACTIONS ON NETWORK SCIENCE AND ENGINEERING}%
{Shell \MakeLowercase{\textit{et al.}}: Bare Advanced Demo of IEEEtran.cls for IEEE Computer Society Journals}

% \IEEEpubid{0000--0000/00\$00.00~\copyright~2015 IEEE}
\IEEEpubid{Copyright (c) 2019 IEEE. Personal use is permitted. For any other purposes, permission must be obtained from the IEEE by emailing pubs-permissions@ieee.org.}
% \IEEEpubid{\makebox[\columnwidth]{\hfill 0000--0000/00/\$00.00~\copyright~2015 IEEE}%
% \hspace{\columnsep}\makebox[\columnwidth]{Published by the IEEE Computer Society\hfill}}


\IEEEtitleabstractindextext{%
\begin{abstract}
%Belief propagation is a \ashkan{powerful inference} technique \sout{to optimize} over probabilistic graphical models, and has been used to solve the community detection problem for networks described by the stochastic block model. In this work, 
We investigate the widely encountered problem of detecting communities in multiplex networks, such as social networks, with an unknown arbitrary heterogeneous structure. 
%being suitable in many real world multiplex networks, 
To improve detectability, we propose a generative model that leverages the multiplicity of a single community in multiple layers,  with no prior assumption on the relation of communities among different layers. 
%and allows a potentially heterogeneous community structure, 
Our model relies on a novel idea of incorporating a large set of generic localized community label constraints across the layers, in conjunction with the celebrated Stochastic Block Model (SBM) in each layer. Accordingly, we build a probabilistic graphical model over the entire multiplex network by treating the constraints as Bayesian priors. We mathematically prove that these constraints/priors promote existence of identical communities across layers without introducing further correlation between individual communities.  The constraints are further tailored to render a sparse graphical model and the numerically efficient Belief Propagation algorithm is subsequently employed.  We further demonstrate by numerical experiments that in the presence of consistent communities between different layers, consistent communities are matched, and the detectability is improved over a single layer. We compare our model with a "correlated model" which exploits the prior knowledge of community correlation between layers. Similar detectability improvement is obtained under such a correlation, even though our model relies on much milder assumptions than the correlated model. 
%When the network has heterogeneous community structures, 
Our model even shows a better detection performance over a certain \ashkan{correlation and signal to noise ratio (SNR)} range. In the absence of community correlation, the correlation model naturally fails, while ours maintains its performance.
\end{abstract}

}

\maketitle

% is not selected <OR> if conference mode is selected - because compsoc
% conference papers position the abstract like regular (non-compsoc)
\IEEEdisplaynontitleabstractindextext

% \fi
\IEEEpeerreviewmaketitle

\section{Introduction}
\label{sec:intro}

A multiplex network structure is a comprehensive representation of real world networks considering that it allows for multiple kinds of relations, and encodes them separately. The multilayer nature of these networks substantially changes their structure and dynamics \cite{Bianconi2013,DeDomenico2013,Cardillo2013,Boccaletti2014,DeDomenico2015a,Mahdizadehaghdam2016} in comparison to single layer representations\cite{Karrer2011,Wang2013}. Despite being one of the main topics of network science for over a decade, the community detection problem has only recently been more closely studied in the context of multilayer networks \cite{PeterJ.Mucha2010,DeDomenico2015a,Loe2015,Wilson2016,Valles-Catala2016,Taylor2016,Stanley2016,Afsarmanesh2016,Paul2017,DeBacco2017}. Community detection in multiplex networks has found numerous applications, such as dynamics \cite{Palla2007} and multi-relation \cite{Szell2010,huang2016consensus} in social networks, evolution of granular force networks \cite{Papadopoulos2016}, and cognitive states of brain networks \cite{Telesford2016}.

It is generally advantageous for community detection to decompose an ordinary network into multiple layers based on additional attributes, and to create a multiplex network for individual layers to potentially unravel entangled structures, such as overlapping communities. It can, however, be difficult to reach this goal with a usually limited knowledge. Carelessly breaking a network into layers can be problematic since it can either retain overlapping communities in a single layer, or decrease detectability of certain communities by breaking them up and distributing them over multiple layers, leading to redundant communities and reduced edge density. This problem arises in many real world multiplex networks. For example, in the networks of protein-genetic interactions, each type of genetic interaction may be used to define a layer, but it is shown to return highly redundant layers, which require recombination \cite{DeDomenico2015}.

\begin{figure}[!t]
\centering
\includegraphics[width=2.5in]{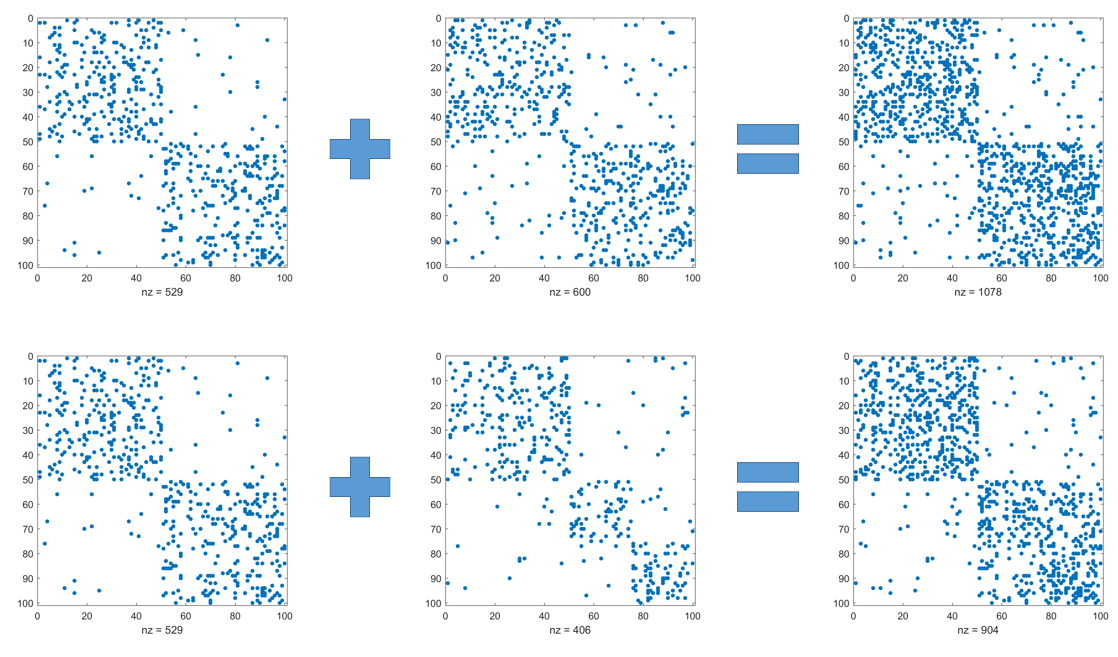}
\caption{Demonstration of potential benefit (upper row) and caveat (lower one) of aggregating multiplex layers. In both rows, left hand side shows the adjacency matrices of two multiplex layers, and right hand side shows that of the aggregated layer by adding the two adjacency matrices. The upper row shows a better community quality of the aggregated network than single layer, while the lower row shows a obscured community structure (notice in the lower row that the consistent community between two layers still gets enhanced quality).}
\label{fig:demo}
\end{figure}

Some recent studies consider the redundancy phenomenon in multiplex networks \cite{DeDomenico2015,Taylor2016,Stanley2016} and try to resolve it by further aggregating the redundant layers. Domenico et al. \cite{DeDomenico2015} utilize tools from quantum information to identify redundant layers and aggregate them hierarchically, thus simplifying the structure. They discovered that many real world multiplex networks, including protein-genetic interactions, social networks, economical and transportation systems, can be significantly simplified by their proposed technique. Taylor et al. \cite{Taylor2016} showed that the detectability of community structure is significantly improved by aggregating layers generated from the same stochastic block model (SBM), which is a popular probabilistic generative model for describing nodes' group memberships \cite{Wang1987}.  Stanley et al. \cite{Stanley2016} proposed a specific multilayer SBM which partitions layers into sets called strata, each described by a single SBM. Layers in a stratum are treated as multiple realizations of the same community structure, thus improving community detection accuracy. A drawback of layer aggregation is that completely consistent community structure between layers is required and needs to be known a priori, otherwise different communities may overlap when aggregated into a single layer, as shown in Fig. \ref{fig:demo}. Domenico et al. \cite{DeDomenico2015a} used the concept of modular flow to show that aggregating layers into a single layer may obscure actual organizations, and that highly overlapping communities exist in some real-world networks. While many algorithms are proposed for overlapping community detection in single layer networks \cite{Palla2005,Esquivel2011,Psorakis2011,Yang2013a,Yang2013b,Nguyen2015,Gamble2015}, the performances remain mediocre due to the loss of layer information.

Inspired by these works, we consider a general multiplex SBM that allows layers to be "partially" redundant, in which case layers may share one or more common communities, and have different ones at the same time (lower row in Fig. \ref{fig:demo}). Our goal is to improve detectability by leveraging the consensus communities without assuming any two layers to belong to the exact same SBM. This not only achieves higher accuracy, but improves detectability of weak consensus communities as well, by combining their information from different layers, which are otherwise too noisy to be detectable individually. Since our model potentially generates a heterogeneous community structure across layers, our method provides a way to detect overlapping communities at theoretically optimal accuracy, when they can be allocated to different layers.

Our method originates from an application of belief propagation algorithm to community detection, as first developed by Decelle et al. \cite{Decelle2011,Decelle2011a}. \NEW{Belief propagation is one algorithms in the Bayesian inference framework, which in turn, is known to yield optimal estimates of communities for a network generated by the underlying SBM \cite{Decelle2011}}. Decelle et al. studied detectability transition, and identified a phase transition point in the parameter space, where all community detection algorithms fail. Since then, some extending works using belief propagation have been reported \cite{Newman2015,Ghasemian2015,Zhang2016,Kawamoto2017}. Ghasemian et al. \cite{Ghasemian2015} extended this method to temporal networks, introducing Dynamic Stochastic Block Model, where nodes gradually change connections and their community memberships over time. While not intended for general temporal networks, our multiplex network model in contrast to \cite{Ghasemian2015}, addresses networks that typically encode multiple relations through layers, and the members of a given community remain unchanged irrespectively of the layer the latter occurs in. Our model also does not enforce a temporal order of the layers as in \cite{Ghasemian2015}. Despite aiming for different types of multiplex networks, a simplified version of \cite{Ghasemian2015} is used as a comparison with our model, in presence of homogeneous and heterogeneous community structures. We show that in different situations, both method show their own strength.

The outline of the paper is as follows. In Sec. \ref{subsec:problem}, we define the problem of community detection and information fusion in multilayer networks. In Sec. \ref{subsec:Bayesian}, we present our proposed stochastic model and a simpler model for comparison. In Sec. \ref{ssec:BP}, we review the belief propagation algorithm and explain the implementation on the proposed model. In Sec. \ref{sec:synthetic}, we show multiple experimental results of the proposed model and discuss its evaluation in detail and its comparison with the simpler model. 

%\begin{figure}[htb]
%	
%	\begin{minipage}[b]{1.0\linewidth}
%		
%		\includegraphics[width=8.5cm]{Fig1_attribute_demo}
%		
%	\end{minipage}
%	
%	\caption{Demonstration of potential benefit (upper row) and caveat (lower one) of aggregating multiplex layers. In both rows, left hand side shows the adjacency matrices of two multiplex layers, and right hand side shows that of the aggregated layer by adding the two adjacency matrices. The upper row shows a better community quality of the aggregated network than single layer, while the lower row shows a obscured community structure (notice in the lower row that the consistent community between two layers still gets enhanced quality).}
%	\label{fig:demo}
%\end{figure}

\section{Problem and method}
\label{sec:problem}

\subsection{Problem Description}
\label{subsec:problem}

%\subsubsection{Problem Statement}
An informed description of our problem of interest is the following: suppose that a multiplex network $W=(V,E(1),E(2),...,E(L))$ is given where $V=\{v_1,v_2,...v_N\}$ is the set of $N$ nodes and $E(l)$ is the set of edges on $V$ at the $l$-th layer. We are to identify a collection $C=\{C_1,C_2,...,C_q\}$ of node communities, where $C_i\subseteq V$ corresponds to a dense subgraph in at least one layer. Although our problem admits overlapping communities, we assume that the co-occurring communities in each layer are disjoint. Each community may also appear in multiple layers, in which case the resulting data multiplicity can be used to improve community detectability by improving the signal to noise ratio (SNR). However, since the occurrence pattern of the communities is not a priorily known, fusing multiple observations of the same community is not straightforward. For a large part of this paper, we assume that the number $q$ of communities is known. However in Section \ref{effect_q}, we briefly discuss the impact of an incorrect choice of $q$ and possible remedies.

%\subsubsection{Mathematical Formulation}

\subsection{Bayesian Solution by Stochastic Modeling}
\label{subsec:Bayesian}
We adopt a Bayesian approach by providing a stochastic generative model  for the observed multiplex network, expressed by a likelihood function $P(W\mid C)$, as well as a prior distribution $P(C)$ on communities. Then, the maximum a-posteriori (MAP) estimate of the communities is obtained by maximizing the a-posteriori distribution, computed according to the Bayes rule:
\[
\hat{C}_\mathrm{ML}=\arg\max\limits_{C}P(C\mid W)=
\arg\max\limits_{C}\frac{P(W\mid C)P(C)}{P(W)},
\]
where $P(W)=\sum\limits_{C^\prime}P(W\mid C^\prime)P(C^\prime)$ is a scaling constant and can be eliminated from optimization. Our generative model utilizes the stochastic block model (SBM), explained in Section \ref{sec:SBM}, which is widely expressed in terms of node-community labeling. For this reason, we provide an alternative representation of the communities by community labeling $T=\{t_i(l)\}$ of the nodes $i$ at different layers $l$. Since, there is a correspondence between possible communities $C$ and the labeling $T$, the generative model $P(W\mid C)$ and the prior $P(C)$ can be equivalently expressed in terms of the labeling as $P(W\mid T)$ and  $P(T)$, respectively. We carefully explain this approach, and the resulting stochastic model is given in Section \ref{sssec:SBMWPP}. We can similarly obtain the MAP estimate $\hat{T}_\mathrm{MAP}$ of $T$ and find its corresponding set of communities, which coincides with $\hat{C}_\mathrm{ML}$, but we resort to a well-known alternative approach, for numerical feasibility. In this approach, we first calculate the marginal probability distribution $p_{i,l}(\alpha)=P(t_i(l)=\alpha\mid W)$ of the labels $\alpha$ of a single node $i$ in a single layer $l$. This is given by
\begin{equation}
P(t_i(l)=\alpha\mid W)=\sum_{T\mid t_i(l)=\alpha}P(T\mid W),
\label{eq:margin}
\end{equation}
where we recall that the posterior distribution $P(T\mid W)$ is calculated by Bayes rule as
\begin{equation}
P(T\mid W)=\frac{P(W\mid T)P(T)}{\sum_{T^\prime}P(W\mid T^\prime) P(T^\prime)}.
\label{eq:global}
\end{equation}
Next, we obtain the maximum marginal a-posteriori probability (MMAP) label estimates $\hat{T}_\mathrm{MMAP}=\{\hat{t}_{i,\mathrm{MMAP}}(l)\}$ by individually maximizing the resulting \textit{posterior marginal} distributions $p_{i,l}(\alpha)$ for every node:
\[
\hat{t}_{i,\mathrm{MMAP}}(l)=\arg\max_\alpha p_{i,l}(\alpha),
\]
from which the corresponding community estimates $\hat{C}_\mathrm{MMAP}$ can be easily obtained. It is shown in \cite{Decelle2011} that $\hat{C}_\mathrm{MMAP}$ is an optimal estimate of the original assignment for large networks with the SBM, which is often slightly better than the MAP estimate $\hat{C}_\mathrm{MAP}$ (ground state) in terms of the number of correct assignments.

Numerical efficiency of the above approach depends on the computation of marginal distributions $p_{i,l}(\alpha)$, which is difficult to perform directly. For example, the denominator in Eq. \eqref{eq:global}, known as the partition function, cannot be exactly calculated unless the system is extremely small or approximate approaches such as Gibbs sampling are used. In Sec. \ref{ssec:BP}, we use a computationally more efficient variational method called belief propagation (BP), which gives the exact marginals $p_{i,l}(\alpha)$ as an approximation of the partition function by a product of marginals, and leads to an efficient implementation of the above approach. We next discuss the generative model in detail.       
%where $W(l)=(V,E(l))$ is the observed network (graph) in the $l^\mathrm{th}$ layer
\subsubsection{\label{sec:SL_SBM}Stochastic Block Model in Single-Layer Network}
\label{sec:SBM}
Stochastic block model (SBM) is commonly used to describe non-overlapping community structures of a single layer network, and plays an important role in our model. Hence, we explain it first. As a generative model, it includes the following parameters: the number of communities $q$, the fraction of the size of each community $\{n_a\}$, the affinity matrix $p=\{p_{ab}\}$ showing the probability of an edge between nodes in communities $a$ and $b$, and the community assignment $t_i\in \{1,...,q\}$ for each node $i$.

A single-layer network is generated from SBM by first assigning to each node one of the community labels $t_i$. The probability of a node being assigned to a community label $a$ is proportional to the size $n_a$ of the community. Then, a pair $(i,j)$ are connected ($A_{ij}=1$ in the adjacency matrix) with probability $p_{t_it_j}$ independently of other pairs. According to the SBM \cite{Wang1987}, if the size of a community is large enough, the community will appear as a block with high probability in the adjacency matrix, under suitable ordering of the nodes.

In benchmark tests, it is common to set $p_{ab}=p_{in}$ if $a=b$, and $p_{ab}=p_{out}$ if $a\neq b$. The constants $p_{in}$ and $p_{out}$ are selected such that the fraction  $\epsilon=p_{out}/p_{in}$ is between 0 to 1, so as to control the community quality in the generated network. $\epsilon=0$ means no connections between two different communities, which represents a high quality community structure. A high $\epsilon$ value \NEW{($\epsilon\approx1$)} means that the connection densities inside and outside the blocks are not significantly distinct, usually reflecting a noisy and weak community structure.

\subsubsection{Generalization to Multiplex Networks}
\label{sssec:SBMWPP}
Now we generalize SBM to multiplex networks. The idea behind our generative model for multiplex networks is that the same community may appear in multiple layers. Each layer $l$ takes a subset of a collection of communities $C$, denoted by $H_l\subseteq C$. If communities $C_a,C_b\in H_l$ and $a\neq b$ (here $a$ and $b$ are community labels), it is required that $C_a \cap C_b= \emptyset$, meaning that overlapping communities are not allowed in any layer. Also, we assume that when a community $C_a$ exists in multiple layers, it refers to the same group of nodes, so that the definition of $C_a$ is independent of the layers. We call these requirements Well Partitioned Property (WPP), and it is an interlayer constraint. WPP has real world relevance a good case being the social network. We can build a multiplex social network using different types of relations, such as contacts, collegial interaction, common interests, etc., in order to disentangle overlapping community structures. However, communities may exist across multiple layers, e.g. a group of close friends may be reflected as the same community in both the rock music network and the soccer fan network. Meanwhile, in these two layers, other people may form inconsistent community structures. In conclusion, we want to build a model, such that only if consistent communities exist between layers, they will be matched and fused.

Under WPP, we may define the community label vector $\bm{t}(l)=(t_1(l),t_2(l),...t_N(l))$ for all $N$ nodes in layer $l$ similarly to the single-layer model in Sec. \ref{sec:SL_SBM}:
\[
t_i(l)=\left\{\begin{array}{cc}
a & i\in C_a,\ C_a\in H_l ,\\
\emptyset & \text{otherwise}.
\end{array}\right.
\]
The community $C_a$ can be easily recovered from the labels by collecting every node labeled by $a$:
\[
C_a=\{v_i\mid\exists l,\ t_i(l)=a\}
\]

The generative model proceeds as follows: the community label vector $\bm{t}(l)$ for nodes in layer $l$ is generated from SBM parameters, under the interlayer constraint WPP. The adjacency matrix of layer $l$ is then independently generated as an ordinary SBM. We propose to formulate the probability of a multiplex network $\{W(l)\}$ and community labels $\{\bm{t}(l)\}$, conditioned on a set of SBM parameters as,\\\\
\textbf{Proposed model}:
\begin{equation}
\begin{aligned}
&P(\{W(l)\},\{\bm{t}(l)\}|p,q,\{n_a\})\\
&=\frac{1}{Z}\prod_{(i,j),(l,l')}f_{check}(t_i(l),t_j(l),t_{i}(l'),t_{j}(l'))\\
&\times \prod_{l=1}^L\left[\prod_{(i,j)\in E(l)}p_{t_i(l),t_j(l)}\prod_{(i,j)\notin E(l)}(1-p_{t_i(l),t_j(l)})\prod_{i}n_{t_i(l)}\right].
\end{aligned}
\label{eq:overall}
\end{equation}

In the following, we break down the formulation and explain each component. We start with a factorized form of the likelihood function, assuming the set of parameters $\theta =\{p,q,\{n_a\}\}$ given,
\begin{equation}
\begin{aligned}
&P(\{W(l)\},\{\bm{t}(l)\}|\theta)\\
=&P(\{W(l)\}|\{\bm{t}(l)\},\theta)P(\{\bm{t}(l)\}|\theta),\\
=&P(\{\bm{t}(l)\}|\theta)\prod_{l=1}^{L}P(W(l)|\bm{t}(l),\theta),\\
\end{aligned}
\label{eq1}
\end{equation}
where $W(l)=(V,E(l)), \ \ l=1,...,L$ is a multiplex layer.

If we look at the product term, $P(W(l)|\bm{t}(l),\theta)$ is the probability that a layer $l$ of the network being generated by a community structure $\bm{t}(l)$. Same as the single-layer SBM, introduced in \cite{Decelle2011},

\begin{equation}
\begin{aligned}
&P(W(l)|\bm{t}(l),\theta)\\
&=\prod_{(i,j)\in E(l)}p_{t_i(l),t_j(l)}\prod_{(i,j)\notin E(l)}(1-p_{t_i(l),t_j(l)}).
\end{aligned}
\label{eq:sbm}
\end{equation}

The other term, $P(\{\bm{t}(l)\}|\theta)$, is the probability distribution over all community patterns satisfying the interlayer constraints from WPP. We express these constraints by a product of local indicator functions $f_{check}$ over the associated community assignment labels $t_i(l)$. Therefore if at least one of the indicator functions is zero (local WPP condition is not satisfied), $P(\{\bm{t}(l)\}|\theta)$ will be zero. Specifically, we propose the distribution of community patterns as:
\begin{equation}
\begin{aligned}
&P(\{\bm{t}(l)\}|\theta)\\
&=\frac{1}{Z}\prod_{i,l}n_{t_i(l)}\prod_{i(l),j(l),i(l'),j(l')}f_{check}(t_i(l),t_j(l),t_i(l'),t_j(l')),
\label{eq:check}
\end{aligned}
\end{equation}
where $Z$ is a suitable normalization constant. The local constraint $f_{check}$ is an indicator function of the state (community label) of the copies of 2 nodes $i,j$ in 2 different layers, $l,l^\prime$ ($i(l)$ means node $i$ in layer $l$). The function $f_{check}$ checks whether the associated labels satisfy WPP, and $f_{check}$ equals one if the following occurs, and is zero otherwise:

\[
\begin{aligned}
&\text{Assume}\ \begin{cases}
t_i(l)=\alpha \\
t_j(l)=\beta
\end{cases},\\
&\text{If}\ \alpha=\beta,\	\text{then}\ \begin{cases}
t_i(l')=t_j(l')=\alpha\\
\text{or}\ \begin{cases}
t_i(l')\neq \alpha\\
t_j(l')\neq \alpha
\end{cases}\\
\end{cases},\\
\\
&\text{If}\ \alpha\neq \beta,\ \text{then}\ \begin{cases}
t_i(l')\neq \beta\\
t_j(l')\neq \alpha
\end{cases}.\\
\end{aligned}
\]

This set of conditions summarize when the labels of two nodes satisfy WPP, as we will discuss in detail next. In practice, given a certain number of communities $q$, we can build a list of all possible combinations that satisfy the above constraint and set $f_{check}=1$. Therefore, the process of evaluating the function $f_{check}$ by verifying the above constraint, can be significantly simplified by storing a look-up table. The look-up table is simple to build for moderate $q$ with a complexity of $q^4$, and only needs to be computed once for a certain $q$ value.

%\AP{I do not like your example here: First, it is not simple for the reader and reviewer (remember when you explain it in the group meetings nobody follows!). Once the reviewer reads your example, she/he will be more confused about your intuition. Always remember that the reviewer reads your paper sentence by sentence and never thinks beyond what you write. So, do not expect them to "understand" you. Second, even if one understands your intuition it does not show why you chose this explicit form, so your explanation is not technical enough. What the reader expects here is a solid explanation that this form is chosen because A, B, .... So, my suggestion is that we write a rigorous theorem. Then, your example can show how the theorem works.}

%Finally, we obtain the overall likelihood of a multiplex network as follows:
%
%\begin{equation}
%\begin{aligned}
%&P(\{W(l)\},\{\bm{t}(l)\}|p,q,\{n_a\})\\
%&=\frac{1}{Z}\prod_{(i,j),(l,l')}f_{check}(t_i(l),t_j(l),t_{i}(l'),t_{j}(l'))\\
%&\times \prod_{l=1}^L\left[\prod_{(i,j)\in E(l)}p_{t_i(l),t_j(l)}\prod_{(i,j)\notin E(l)}(1-p_{t_i(l),t_j(l)})\prod_{i}n_{t_i(l)}\right].
%\end{aligned}
%\label{eq:overall}
%\end{equation}

\subsubsection{Characterizing WPP}

%We are able to show that a multi-layer community structure $(C,\{H_l\})$ satisfies WPP if and only if the $f_{check}$ values for every pair of nodes and every two layers equal one (See Appendix \ref{WPP}). 
We are able to proof that a multi-layer community structure $(C,\{H_l\})$ satisfies WPP, if and only if, for the labels of every pair of nodes and every two layers, the value of function $f_{check}$ equals one and hence $P(\{\mathbf{t}(l)\}\mid\theta)=1$. The general proof is in the appendix.

Here we show a simple example to demonstrate one of the constraints. %More examples are provided in appendix.
\begin{figure}[htb]	
	\begin{minipage}[b]{1.0\linewidth}
		
		\includegraphics[width=8.5cm]{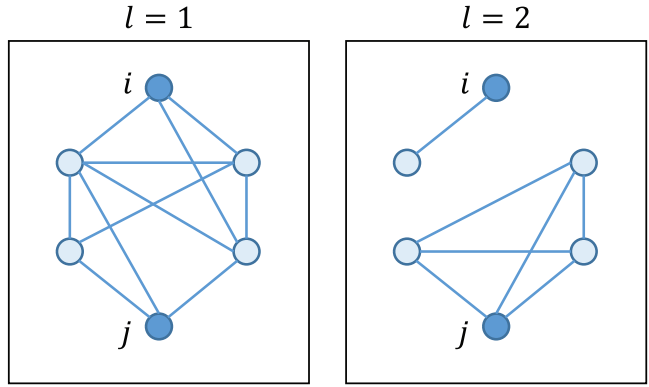}
		
	\end{minipage}
	
	\caption{An example of a two-layer network with different community structures.}
	\label{fig:consexa}
\end{figure}

Fig. \ref{fig:consexa} shows a situation where community structures in two layers are different (each connected component in a layer is a community). According to the connectivity patterns, we observe that $t_i(1)=t_j(1)$ and $t_i(2)\neq t_j(2)$ as $i,j$ are in the same community in layer 1, while they are in different communities in the second layer.  We conclude that, in the second layer, neither node $i$, nor node $j$ can be assigned to the same community as the one in the first layer, and hence at least 3 communities are required for a correct assignment. This simple intuition is reflected in the definition of $f_{check}$ (the case of $\alpha=\beta$), where $l=1,l^\prime=2$.

\begin{figure}[htb]
	\begin{minipage}[b]{1.0\linewidth}
		\includegraphics[width=8.5cm]{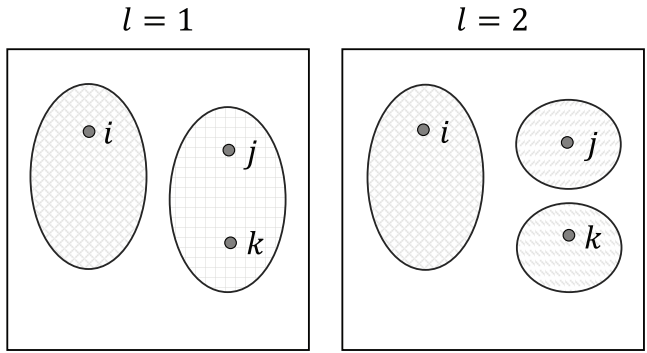}
		
	\end{minipage}
	
	\caption{An example of a two-layer network with partially consistent and partially different community structures. Each ellipse represents a community.}
	\label{fig:consexa2}
\end{figure}

We observe that the constraints in $f_{check}$, when utilized in a Bayesian learning algorithm, ensure that distinct communities in different layers will not be assigned the same label and not be confounded as one community, so that the structural information will not be mixed up and obscured. This is, according to our example, due to the fact that assigning the same labels to unequal communities will lead to violation of constraints, and make corresponding $f_{check}$ functions zero and consequently a zero-value posterior distribution $P(\{W(l)\},\{\bm{t}(l)\}|\theta)$. Another role of the constraints is to equally assign consistent communities in different layers, and fuse the structural information to improve detectability. This is illustrated in our example, depicted in
Fig. \ref{fig:consexa2}, where only the community for node $i$ is consistent between two layers, and our goal is to assign to the copies of node $i$ the same community label. Notice that in total, 4 communities are involved in this example. If we set $q=4$, any community assignment with $t_i(1)\neq t_i(2)$ will violate the $f_{check}$ constraints, which in turn will force $t_i(1)=t_i(2)$ in the Bayesian learning algorithm. For example, let $t_i(1)=a_1$ and $t_i(2)=a_2$. Since $t_i(1)\neq t_j(1)$, according to the constraint where $\alpha \neq \beta $, we know that $t_i(1)\neq t_j(2)$ and $t_i(2)\neq t_j(1)$, and we let $t_j(1)=a_3$ and $t_j(2)=a_4$, and therefore $t_k(1)=a_3$. Similarly using the same constraint, we know $t_i(1)\neq t_k(2)$. We derive that $t_k(2)\neq a_1$, and due to the community structure in layer 2, $t_k(2)\neq a_2,a_4$. Then again using the constraint of $\alpha \neq \beta$ on $t_j(2)\neq t_k(2)$, we derive that $t_k(2)\neq t_j(1)=a_3$. We find out that $t_k(2)$ is not able to choose from any of the four community labels without violating the constraints. However if we set $q=5$, we can find community assignments with $t_i(1)\neq t_i(2)$ while satisfying the constraints (for example $t_i(1)=a_1, t_j(1)=t_k(1)=a_2, t_i(2)=a_3, t_j(2)=a_4, t_k(2)=a_5$), in which case, the communities for node $i$ in the two layers will be independently treated and detectability cannot be improved. This also demonstrates the important role of the number $q$ of communities as a design parameter.  Although we may not know \textit{a priori} the actual number of communities, this number can be estimated \cite{Decelle2011}. We will discuss later (in Sec. \ref{effect_q}) how the number of communities affects detection results.

%Notice here that we implicitly assume the number of communities to be three. In fact, by limiting the number of communities to the actual value, same communities in different layers will be forced to have the same labels. How the number of communities affects detectability will be discussed later. \AP{This is very vague for the reader. We do not expect you to explain such details here. Maybe you can discuss this in the results part.}

%We will have a factor for each combination of 4 nodes to check the constraints, which will be a huge number. However we expect these factors are largely redundant and we only need to sample a small fraction to avoid overlapping communities in each layer. In other words, we can greatly reduce the number of messages in this part. (maybe refer to ECC)

\subsubsection{A Prototypical Multiplex Model}
\label{sssec:oracle}

To discuss the performance of our proposed approach in Sec. \ref{sssec:BPconstraint}, we present a simpler "correlated model" without overlapping communities, but with variable and correlated ones in different layers. The correlated model is similar to the DSBM (Dynamic Stochastic Black Model) introduced in \cite{Ghasemian2015}. This model achieves the best performance when the communities in different layers are the same, since layer consistency is used as prior knowledge, much like the layer aggregation method in \cite{Taylor2016}. However, the presence of such a strong prior information is not always realistic, and this model only serves as an oracle bound for our proposed model as in Eq. \eqref{eq:overall}.

We now modify the above SBM model to a correlated multilayer structure, following the same Bayesian description as in Eqs. \eqref{eq1} and \eqref{eq:sbm}, nevertheless different from our model in Eq. \eqref{eq:overall}, in that the community assignment prior is instead given by:
\begin{equation}
\begin{aligned}
P(\{\bm{t}(l)\}|\theta)=\prod_{i(l),i(l')}f(t_i(l),t_i(l')),
\end{aligned}
\end{equation}
where $f(\cdot)$ is a factor function for the correlation of community assignment of the same node $i$ in two layers, indicating the probability of different $(t_i(l),t_i(l'))$ combinations:
\[
f(t_i(l),t_i(l'))=
\begin{cases}
p_{same},\  t_i(l)=t_i(l')\\
1-p_{same},\  t_i(l)\neq t_i(l')
\end{cases},
\]
where $p_{same}\in [0,1]$ is the probability of consistent community labels between the same node in two layers. In a special case, if we constrain the number of both layers and communities to 2, when $p_{same}>0.5$, node labels between two layers are correlated, when $p_{same}<0.5$, anti-correlated, and when $p_{same}=0.5$, uncorrelated. Note that when $p_{same}<1$, it allows the same community label to correspond to different sets of nodes in different layers. For $q>2$, $p_{same}=0.5$ is still the threshold above which the communities become correlated, but then $f(t_i(l),t_i(l'))$ needs to be normalized to be the real probability. Similarly to Eq. \eqref{eq:overall}, we propose the following Bayesian model:

\begin{equation}
\begin{aligned}
&P(\{W(l)\},\{\bm{t}(l)\}|p,q,\{n_a\})\\
&=\frac{1}{Z}\prod_{i(l),i(l')}f(t_i(l),t_i(l'))\\
&\times \prod_{l=1}^L\left[\prod_{(i,j)\in E(l)}p_{t_i(l),t_j(l)}\prod_{(i,j)\notin E(l)}(1-p_{t_i(l),t_j(l)})\prod_{i}n_{t_i(l)}\right].
\end{aligned}
\label{eq:oracle}
\end{equation}

Unlike WPP, this model assumes variable communities and correlation between community assignments of a single node between layers. This may be too ideal relative to Eq. \eqref{eq:overall}, since it adds to the model some privileged prior knowledge which is uncommon in real scenarios. We will later compare the model in Eq. \eqref{eq:oracle} with the constrained multiplex model proposed in Eq. \eqref{eq:overall}.

\subsection{BP algorithm for multilayer community detection}
\label{ssec:BP}

Belief Propagation is an efficient message-passing method for inference problems. Message-passing appears in various contexts, and with various references, such as sum-product algorithm, belief propagation, Kalman filter and cavity method which is used to compute phase diagrams of spin glass systems. Yedidia et al. \cite{Yedidia2002,Yedidia2005} gave a detailed introduction to Belief Propagation and its connection to free energy.

We use the BP algorithm for calculating the marginal posterior distributions $p_{i,l}(\alpha)$ as explained in Section \ref{subsec:Bayesian}. To that end, we will represent our model in Eq. \eqref{eq:overall} as a factor graph. A factor graph is composed of factor nodes and variable nodes. Each variable node corresponds to an actual node in our multiplex network. A factor node corresponds to a factor in Eq. \eqref{eq:overall}. In a tree-like Bayesian network, each factor can also be interpreted as a conditional probability distribution $p(x_i|Parent(x_i))$. Here $x_i$ corresponds to a variable node and $Parent (x_i)$ denotes its parent nodes \cite{Yedidia2002}. A factor node is connected to its contributing variables, therefore connecting a variable node and all its parent variable nodes. As seen in Eq. \eqref{eq:overall}, two types of factor nodes arise in our case: constraint ($f_{check}$) nodes, connected to four variable nodes, and the remaining SBM nodes, connected to two variables (See Fig. \ref{fig:factorgraph}).

In BP, "messages" are reciprocally sent between variable nodes and factor nodes. These messages are a set of equations about the estimates of the conditional marginals. These equations are self-consistent in the sense that they will converge to a consistent solution upon repeatedly iterating. On factor graphs, messages $m^{i\rightarrow a}$ from variable nodes $i$ to factor nodes $a$ are different from the reversed ones $m^{a\rightarrow i}$ and are given by:
\begin{equation}
\begin{aligned}
&m^{i\rightarrow a}(x_i):=\prod_{c\in N(i)\backslash a} m^{c\rightarrow i}(x_i)\\
&m^{a\rightarrow i}(x_i):=\sum_{\textbf{x}_a\backslash x_i}f_a(\textbf{x}_a)\prod_{j\in N(a)\backslash i}m^{j\rightarrow a(x_j)},
\end{aligned}
\label{eq:variablefactor}
\end{equation}
where $N(i)\backslash a$ denotes the neighbors of the variable node $i$ except $a$, and $\textbf{x}_a\backslash x_i$ denotes the neighbors of the factor node $a$ except node $i$. Basically, a variable-to-factor message is proportional to the product of all other incoming messages to the variable node, while a factor-to-variable message is the posterior marginal distribution of the variable based on the individual factor, and assuming other incoming messages to the factor as independent priors.

The computational complexity of BP is low. To obtain a marginal probability distribution of an objective node in graphs with no loops, one starts from all the leaves and uses all messages only once, toward the objective node. In practice, one starts with random initial messages, and let them update iteratively, until they converge to a fixed point, or until they meet a stopping criterion. Hence, for a fixed number of iterations, the computation time is $O(|E|)$. In a generated sparse graph where we fix the average degree, the computation time is $O(N)$ . After convergence, the marginal distribution (also called belief) of a node can be calculated using all incoming messages:

\begin{equation}
b^{i}(x_i)\propto \prod_{c\in N(i)} m^{c\rightarrow i}(x_i).
\label{eq:BPmarginal}
\end{equation}

While, in the presence of cycles, messages may theoretically require infinite iterations to converge, BP has been observed to perform well in graphs that are locally tree-like even if they have many loops\cite{Decelle2011}. Notice that in loopy graphs, the order of message passing is arbitrary and often heuristic.

\subsubsection{Message Passing for Single-Layer SBM}

For single layer networks, ordinary SBM is used to describe community structures. In \cite{Decelle2011} it is shown that since each factor is exactly connected to two variables, the two steps in Eq. \eqref{eq:variablefactor} can be combined to yield a single node-to-node message passing step as follows:
%The likelihood of a network is shown in Eq. \eqref{eq:sbm}, and
%\AP{remove(The corresponding message update equations \cite{Decelle2011} are)}
\[
m_{t_i}^{i\rightarrow j}=\frac{1}{Z^{i\rightarrow j}}n_{t_i}\prod_{k\in N(i)\backslash j}\left[\sum_{t_k}c_{t_kt_i}^{A_{ik}}(1-\frac{c_{t_kt_i}}{N})^{1-A_{ik}}m_{t_k}^{k\rightarrow i}\right],
\]
where $Z^{i\rightarrow j}$ is a normalization constant. $n_{t_i}$ is the fraction of the size of the community $t_i$ (assigned to node $i$), which represents local evidence for node $i$. $c_{t_kt_i}$ is the rescaled connection probability between nodes in communities $t_k$ and $t_i$ respectively i.e., $c_{ab}=Np_{ab}$ ($N$ is the number of nodes). Finally, $A_{ik}$ is an element of the  adjacency matrix of the network.

Decelle et al. \cite{Decelle2011} further use the following mean field approximation to simplify the influence from unconnected nodes,

\begin{equation}
m_{t_i}^{i\rightarrow j}=\frac{1}{Z^{i\rightarrow j}}n_{t_i}e^{-h_{t_i}}\prod_{k\in N(i)\backslash j}\left(\sum_{t_k}c_{t_kt_i}m_{t_k}^{k\rightarrow i}\right),
\label{eq:BPforCD}
\end{equation}
where $h$ is an external field, and expressed as,

\[
h_{t_i}=\frac{1}{N}\sum_k \sum_{t_k}c_{t_k t_i}b^k_{t_k}.
\]

Here $b^k_{t_k}$ is the belief at node $k$ for community label $t_k$, corresponding to our objective in Eq. \eqref{eq:margin}. The belief at node $i$ is written as,

\[
b^i_{t_i}=\frac{1}{Z^{i}}n_{t_i}e^{-h_{t_i}}\prod_{k\in N(i)}\left(\sum_{t_k}c_{t_kt_i}m_{t_k}^{k\rightarrow i}\right).
\]

Clearly, Eq. \eqref{eq:BPforCD} bears a similar structure to a combination of the two steps in Eq. \eqref{eq:variablefactor}. Note that the inner summation part in parentheses in Eq. \eqref{eq:BPforCD} is in the form of a message from a factor node to a variable node i.e., the second line in Eq. \eqref{eq:variablefactor}, while the outside product manifests message passing in the first line of Eq. \eqref{eq:variablefactor} from a variable node to a factor node. We observe that the message \eqref{eq:BPforCD}, being from variable $i$ to variable $j$, essentially bypasses the factor node lying between these two variable nodes, hence further reducing complexity.

\subsubsection{Multiplex Network as a Message Passing Model}
\label{sssec:BPconstraint}

Since the interlayer constraint function in Eq. \eqref{eq:check} is defined by 4 variable nodes rather than pairwise interaction, we can no longer combine the two messages in Eq. \eqref{eq:variablefactor} and directly write inter-layer messages between variable nodes. We instead opt to explicitly write inter-layer messages from variable nodes to factor nodes. For the sake of consistency, we do the same for intralayer messages. The factor graph is illustrated in Fig. \ref{fig:factorgraph}. The message update equations are shown below.
% \begin{figure}
\begin{strip}
% 	\begin{minipage}{1.0\linewidth}
		\textbf{Proposed update equations}:
		
		Intra-layer message:
		
		\begin{equation}
		\begin{aligned}
		&m_{t_i}^{i\rightarrow a}(l)=\frac{1}{Z^{i\rightarrow a }(l)}n_{t_i(l)}e^{-h_{t_i(l)}}\prod_{d\in N_{intra}(i(l))\backslash a}\left(\sum_{t_d}c_{t_dt_i(l)}m_{t_d}^{d\rightarrow i(l)}\right)\\
		& \times \prod_{c\in N_{inter}(i)}\left(\sum_{t_j(l),t_i(l'),t_j(l')} f_{check}(t_i(l),t_i(l'),t_j(l),t_j(l'))\prod_{k\in N_{inter}(c)\backslash i} m_{t_k}^{k\rightarrow c}\right).
		\end{aligned}	
		\label{BPcomplete}
		\end{equation}

		Inter-layer message:
		
		\begin{equation}
		\begin{aligned}
		&m_{t_i}^{i\rightarrow c}=\frac{1}{Z^{i\rightarrow c }}n_{t_i}e^{-h_{t_i(l)}}\prod_{d\in N_{intra}(i)}\left(\sum_{t_d}c_{t_dt_i}m_{t_d}^{d\rightarrow i}\right) \times\\
		& \prod_{c^*\in N_{inter}(i)\backslash c}\left(\sum_{t_j(l),t_i(l'),t_j(l')} f_{check}(t_i(l),t_i(l'),t_j(l),t_j(l'))\prod_{k\in N_{inter}(c^*)\backslash i} m_{t_k}^{k\rightarrow c^*}\right),
		\end{aligned}	
		\label{BPcomplete2}
		\end{equation}	
		
		where $N_{intra}(i)$ represents intra-layer neighbors of node $i$, and $N_{inter}(i)$ the inter-layer ones (the constraint-checking factors). \NEW{$h_{t_i(l)}$ is the external field in layer $l$, referring to the single layer version in Eq. \eqref{eq:BPforCD}.}
		
		To calculate the node belief:
		\begin{equation}
		\begin{aligned}
		&b_{t_i}^{i}(l)=\frac{1}{Z^{i}(l)}n_{t_i(l)}e^{-h_{t_i(l)}}\prod_{d\in N_{intra}(i(l))}\left(\sum_{t_d}c_{t_dt_i(l)}m_{t_d}^{d\rightarrow i(l)}\right)\times\\
		&  \prod_{c\in N_{inter}(i(l))}\left(\sum_{t_j(l),t_i(l'),t_j(l')} f_{check}(t_i(l),t_i(l'),t_j(l),t_j(l'))\prod_{k\in N_{inter}(c)\backslash i(l)} m_{t_k}^{k\rightarrow c}\right).
		\end{aligned}	
		\label{BPcomplete_belief}
		\end{equation}
% 	\end{minipage}
\end{strip}
% \end{figure}

\begin{figure}[tb!]
	
	\begin{minipage}{1.0\linewidth}
		
		\includegraphics[width=8.5cm]{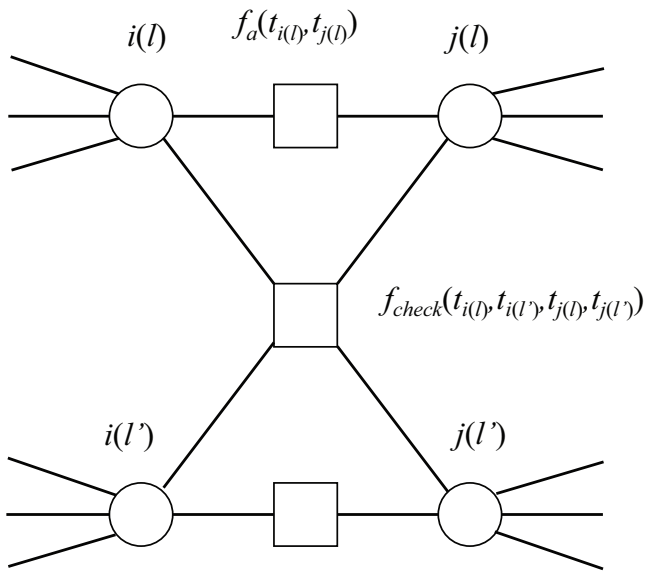}
		
	\end{minipage}
	
	\caption{An example of a factor graph for our model. A circle is a variable node, and a square is a factor node. \NEW{There are two types of factor nodes. One is within each layer, $f_a(t_i(l),t_j(l))=c_{t_i(l),t_j(l)}$, representing the likelihood of node $i$ having label $t_i(l)$ and node $j$ having label $t_j(l)$}. Another is between layers, $f_{check}(t_i(l),t_i(l'),t_j(l),t_j(l'))$, representing the local constraints of community labels.}
	\label{fig:factorgraph}
\end{figure}

Note that the marginal posteriors are given by the beliefs as $p_{il}(\alpha)=b^i_{t_i(l)=\alpha}(l)$. For experimental purposes, and clarity, we write down the belief propagation equations for a two layer network, similarly for the following model. The associated resulting message passing algorithm is shown below as a pseudocode. The "for" loops, which update the messages, can be easily executed in a parallel or distributed fashion for large networks. In our experiments, a serial version of the algorithm is implemented. In each step, one edge is randomly selected without replacement and the corresponding message is updated, which influences the following updates of other edges.

\begin{algorithm}[t]
	\caption{BP for constrained multiplex networks}\label{alg:euclid}
	\begin{algorithmic}[1]
		\State Initialize belief vector for each node in each layer
		\State Compute initial messages and field $h$(more detail)
		\While{$t<t_{max}$ and conv$>$criterium}%\Comment{We have the answer if r is 0}
		\State conv=0; t=t+1
		\For{layer $l$ from 1 to $L$}
		\For{every directed edge $i\rightarrow j$ in layer $l$}
		\State Update message $m^{i\rightarrow j}(l)$ according to Eq. \eqref{BPcomplete}
		\State Update message $m^{i\rightarrow c}$ according to Eq. \eqref{BPcomplete2}
		\EndFor
		\For{every node $i$ in layer $l$}
		\State Update belief $b^{i}(l)$ according to Eq. \eqref{BPcomplete_belief}
		\EndFor
		\State Update field $h(l)$ in layer $l$
		\EndFor
		\State conv$=\sum|m_{new}-m_{old}|$
		\For{every ordered pair of layers $l$ and $l'$}
		\For{every ordered pair of nodes $i$ and $j$}
		\State Update message from $i$ in layer $l$ to the constraint factor node between $i$ and $j$, $m^{i\rightarrow c}(l)$ according to Eq. \eqref{BPcomplete2}
		\EndFor
		\EndFor
		\EndWhile
		\State Compute group assignment
		\State Compute accuracy
		\label{BP}
	\end{algorithmic}
\end{algorithm}

\subsubsection{An Oracle Limit: Correlated Variable Communities}
\label{sssec:BPcorr}

\begin{figure}[htb!]
	
	\includegraphics[width=5.5cm]{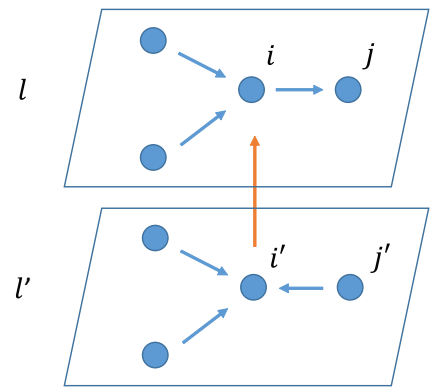}
	
	\caption{The red arrow shows the interlayer message from $i(l')$ to $i(l)$.}
	\label{fig:interlayer}
\end{figure}

We now show the message passing expression for the correlated-community model in Eq. \eqref{eq:oracle}. The message paths are illustrated in Fig. \ref{fig:interlayer}, highlighting inter-layer messages and intra-layer ones. Since every factor node connects only two variable nodes, we can bypass the factor nodes and write messages between variable nodes as in the figure.

\begin{floatEq}
% \begin{strip}
	\begin{minipage}{1.0\linewidth}
		\textbf{Proposed update equations}:
		
		Intra-layer message:
		\[
		m_{t_i}^{i\rightarrow j}(l)=\frac{1}{Z^{i\rightarrow j}(l)}n_{t_i(l)}e^{-h_{t_{i}(l)}}\prod_{k\in N(i(l))\backslash j(l)}\left(\sum_{t_k}c_{t_kt_{i}}(l)m_{t_k}^{k\rightarrow i}(l)\right)\times \sum_{t_i(l')}f(t_i(l),t_i(l'))m^{i(l')\rightarrow i(l)}_{t_i(l')},
		\]
		Inter-layer message:
		\[
		m_{t_i(l')}^{i(l')\rightarrow i(l)}=\frac{1}{Z^{i(l')\rightarrow i(l)}}n_{t_i(l')}e^{-h_{t_i(l')}}\prod_{k\in N(i(l'))}\left(\sum_{t_k}c_{t_kt_i}(l^\prime)m_{t_k}^{k\rightarrow i}(l^\prime)\right).
		\]
	\end{minipage}
% \end{strip}
\end{floatEq}

\section{Detectability transition of constrained multiplex networks}
\label{sec:synthetic}

%\subsubsection{Evaluation metrics}

\subsection{Homogeneous multiplex network}

%\AP{(all experimental details should move to the next section. remove: When we test this on a two-layer network with both layer from the same SBM model (two 100-node blocks), we are providing perfectly consistent communities between two layers. We vary the value of factor function by letting:) In particular, we consider the following correlation factors in our experiments:}

In this section, we report the results of the Bayesian method in Section \ref{subsec:Bayesian} with the message-passing algorithms developed in Sec. \ref{sssec:BPconstraint} and \ref{sssec:BPcorr}. We set the experimental scenario to consist of a 2-layer network with 200 nodes, where each layer is randomly generated according to a SBM. The nodes are partitioned into two communities of equal size, which are present in both layers. This is a result of the probability having the same labels between two layers, i.e. $p_{same}=0$ or 1 in the correlated model, all the while simultaneously satisfying the WPP. For each algorithm, we observe community detectability transition by varying $\epsilon=p_{out}/p_{in}$ in the SBM. The transition is quantitatively characterized by a normalized agreement score $Q\in [0,1]$ (referring to "agreement" in \cite{Decelle2011}),

\[
Q(\{t_i^*(l)\},\{\hat{t}_i(l)\})=\max_\pi \frac{\frac{1}{N}\sum_i{\delta_{t_i^*(l),\pi(\hat{t}_i(l))}-\max_an_a}}{1-\max_an_a},
\]
where $\{t_i^*(l)\}$ is the ground truth community labels, $\pi$ is one of the permutations of estimated community labels $\{\hat{t}_i(l)\}$, and $\max_an_a$ is the size of the largest community. $\frac{1}{N}\sum_i{\delta_{t_i^*(l),\pi(\hat{t}_i(l))}}$ is called agreement score and represents the overlap between estimated community labels and ground truth.

\begin{figure}[htb]
	
	\begin{minipage}[b]{1.0\linewidth}
		
		\includegraphics[width=8.5cm]{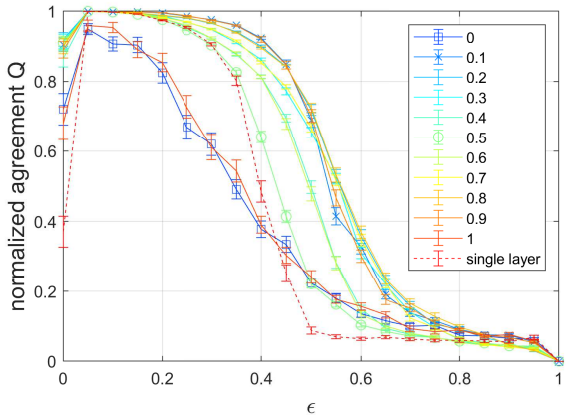}
		
	\end{minipage}
	
	\caption{Detectability transition curves for various $p_{same}$ ranging from 0 to 1. The slower the normalized agreement score $Q$ \cite{Decelle2011} drops down, the better the detectability. Data points for $p_{same}=0,0.1,0.5,0.9,1$ are connected by solid lines. The results are averaged over 100 experiments. The error bars represent standard errors.}
	\label{fig:diff_factors}
\end{figure}

For the correlated-community model in Sec. \ref{sssec:oracle}, we observe transitions curves under various value of $p_{same}$ in the algorithm in Sec. \ref{sssec:BPcorr}. In Fig. \ref{fig:diff_factors}, for $p_{same}=0.5$, detectability transition is similar to that in a single layer (red dash line) \cite{Decelle2011}, because we are practically treating them as independent layers. Except for $p_{same}=0$ or 1, high correlation (such as $p_{same}=0.9$) or anti-correlation (such as $p_{same}=0.1$) between labels increases detectability significantly. We conjecture that the poor performance for $p_{same}=0$ or 1 is due to its low tolerance of wrong intermediate label, leading to a lower chance of convergence to the correct fixed point. The fluctuation of the $p_{same}=0$ or 1 curves also indicates that the convergence is not stable in these cases, especially considering the loopy factor graph.

This naive assumption that all nodes in different layers have correlated community labels is, however, the same as directly connecting corresponding nodes in two layers without any further weight adjustment over messages. \NEW{In this case, all nodes in each layer are assumed as uniformly correlated. This assumption from the correlated model is reasonable for certain types of multiplex networks such as temporal networks. However, to account for heterogeneous structure, and a more realistic case of unknown prior knowledge of consistent communities, it will be more suitable to use our generative model with label constraint.}

\begin{figure}[htb]
	
	\begin{minipage}[b]{1.0\linewidth}
		
		\includegraphics[width=8.5cm]{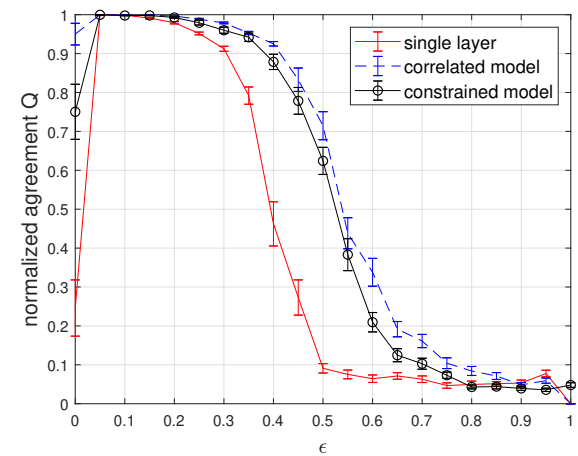}
		
	\end{minipage}
	
	\caption{Detectability transition curves for a single layer network, a correlated double layer network and a double layer network with constraint. \NEW{The synthetic network is a two-layer network, where each layer is independently generated by the same SBM model such that 200 nodes form two equal communities. Nodes have the same community labels in layer 1 and 2, while the edges are different across layers, so information between two layers can be easily combined. $p_{same}$ is 0.9, only used as the inference parameter for the correlated model.} The results are averaged over 30 experiments. Error bars represent the standard error of the experiments.}
	\label{fig:singleVScorrVSconstr}
\end{figure}

\NEW{Fig. \ref{fig:diff_factors} shows that a two-layer network is enough to exhibit the strength of the correlated model. To directly compare the constrained multiplex model in Sec. \ref{sssec:SBMWPP} with the correlated one in Sec. \ref{sssec:oracle}, we follow the same experimental setting as in Fig. \ref{fig:diff_factors}, and test both methods on the homogeneous double layer network. We make sure that each layer has the same community structure and is independently generated by the same SBM parameters: 200 nodes which are divided into two equal communities. Note that we do not generate the network from the correlated model, although the correlated model fits it.} We vary $\epsilon=p_{out}/p_{in}$ to observe the detectability transitions. The result is shown in Fig. \ref{fig:singleVScorrVSconstr}, where we include the transition curve for a single layer (red line) as a reference. Similarly to the correlated model (blue line), the constrained model (black line with circle marks) fails around similar $\epsilon$ values. They both perform much better (fail for larger $\epsilon$) than a single layer.

Note that in the correlated model, we know \textit{a priori} that the community labels are correlated between two layers. In the constrained model, we, however, do not specifically have that prior knowledge. Just by enforcing WPP constraints and limiting the number of communities to 2, we can still achieve a similar performance improvement. This is beneficial for real world networks, since in practice we often have limited prior information about consistent communities. \NEW{Indeed, in this experiment, this prior knowledge may also be inferred in the correlated model, setting interlayer correlation as a parameter and using the EM algorithm \cite{Dempster1977}. However, in more complex cases where, for example, community structure in two layers can not be simply described by a single correlation parameter, the correlated model will face difficulty, as we will show in the next section.}

One may suspect that as long as the blocks are consistent, the detectability can be automatically improved regardless of such correlation being available to the model. This is clearly not the case for the correlated model as in Fig. \ref{fig:diff_factors}, since setting $p_{same}=0.5$, does not include correlation in the model, and the performance is poorer and similar to a single layer setting.

\subsection{Heterogeneous multiplex network}\label{hete_exp}
\begin{figure}[htb]
	
	\begin{minipage}[b]{1.0\linewidth}
		
		\includegraphics[width=8.5cm]{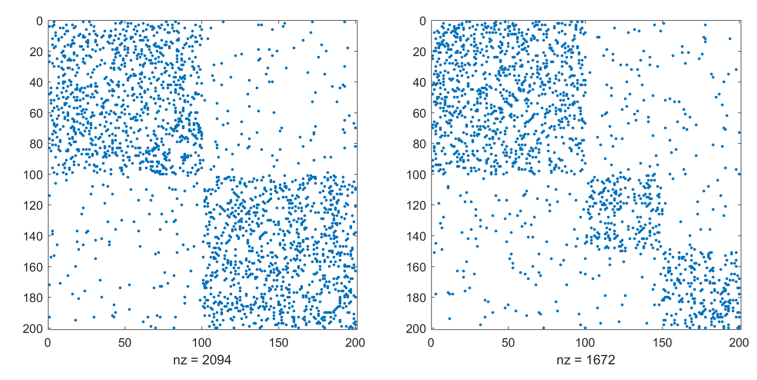}
		
	\end{minipage}
	
	\caption{An example ($\epsilon=0.2$) of the heterogeneous network generated to test the constraint multiplex model. There are in total four distinct communities. First 100 nodes in two layers form the same community, while the rest 100 nodes form three different communities in two layers.}
	\label{fig:hete}
\end{figure}

The constrained model being the only model that naturally generates heterogeneous networks, shows the advantage over the correlated model or single layer networks. In the following we compare the community detection performance between the constrained model and the correlated model on heterogeneous networks. \NEW{We construct a double layer network of 200 nodes, with $\epsilon\in[0,1]$ An example of the synthetic network is shown in Fig. \ref{fig:hete}. In the first layer, the first 100 nodes form a community and the remaining 100 nodes are assigned to another community. In the second layer the first 100 nodes still form a community but the remaining 100 nodes are divided into two equal communities.} By limiting the total number of communities to $q=4$, we expect the belief of the first 100 nodes in both layers to converge to the same label, and the remaining 100 nodes in two layers to converge to three different labels (refer to Fig. \ref{fig:hete}). We refer to this expected result as WPP-satisfying labels and other results as error.

\begin{figure}[htb]
	
	\begin{minipage}[b]{1.0\linewidth}
		
		\includegraphics[width=8.5cm]{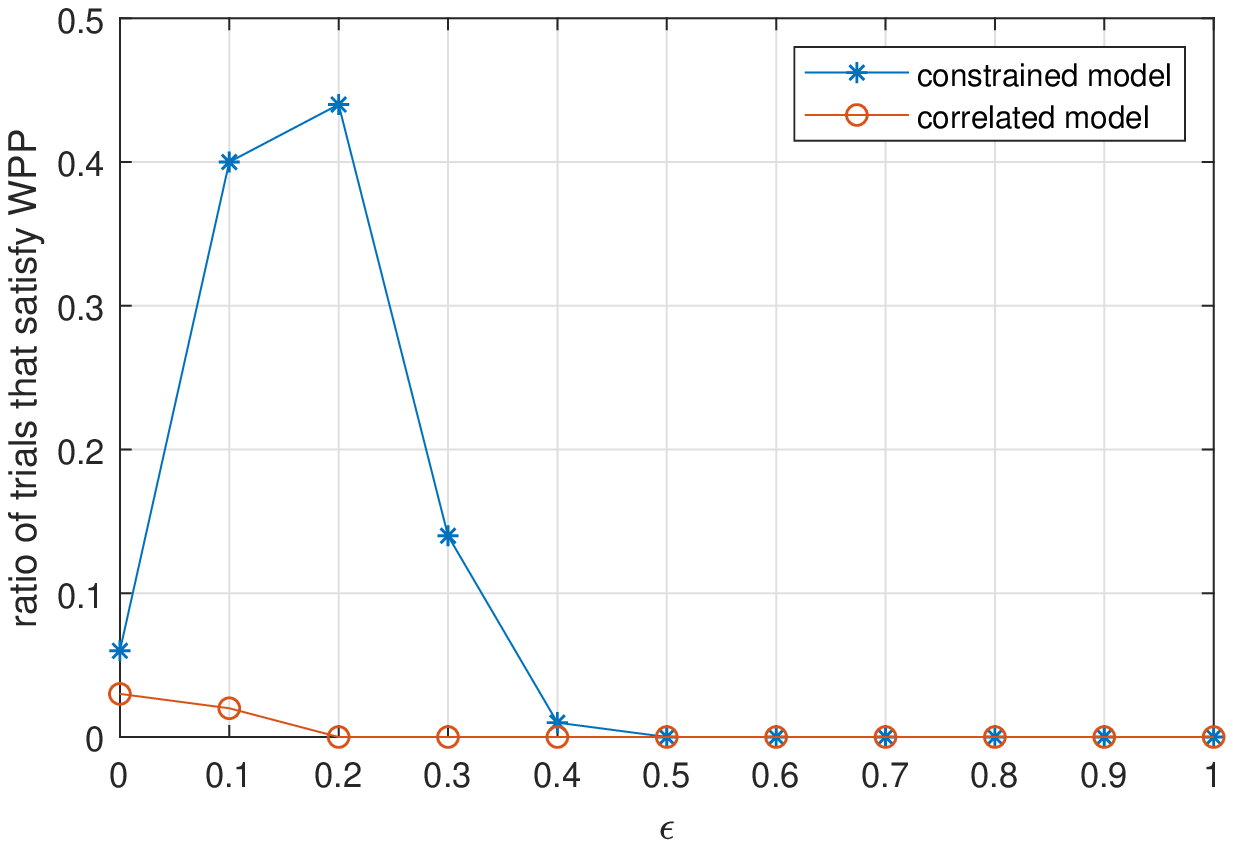}
		
	\end{minipage}
	
	\caption{Fraction of trials that result in WPP-satisfying labels for different $\epsilon$. For each of the five clusters of nodes in two layers, we determine the label by the majority of node labels in that cluster. Then we check if these five labels satisfy WPP. Refer to Fig. \ref{fig:hete} for synthetic network setup.}
	\label{fig:ratio}
\end{figure}

We performed 100 independent trials of tests using both models, and count the fraction of the tests that result in WPP-satisfying labels. As in Fig. \ref{fig:ratio}, when $\epsilon\in [0,0.4]$, our constrained model yields WPP-satisfying labels in some of the trials, while the correlated model is able to achieve that only for $\epsilon\in [0,0.1]$. Also, the constrained model has significantly higher likelihood to yield correct labels, i.e., has the messages converge to the correct point, when $\epsilon\in [0,0.4]$. Note that in this experiment, for each layer, we do not limit the number of communities to the correct value (i.e. two communities for layer 1 and three for layer 2), which means each node in a layer will freely choose from 4 different labels. If we detect communities independently in two layers, which corresponds to setting no constraint, the chance of WPP-satisfying labels is no more than $4!/(P^4_2\times P^4_3)=1/12$, where $P^n_k$ is $k$-permutation of $n$. Our result does show an advantage in identifying consistent communities in heterogeneous networks, while the correlated model is unsuitable for this task. The detection error may be attributed to local minima which violate the constraint (WPP) to some degree, with, however, sufficient resilience for the messages to converge. In practice, we can run the algorithm multiple times and choose the results that more likely converged to a correct point.

\begin{figure}[htb]
	
	\begin{minipage}[b]{1.0\linewidth}
		
		\includegraphics[width=8.5cm]{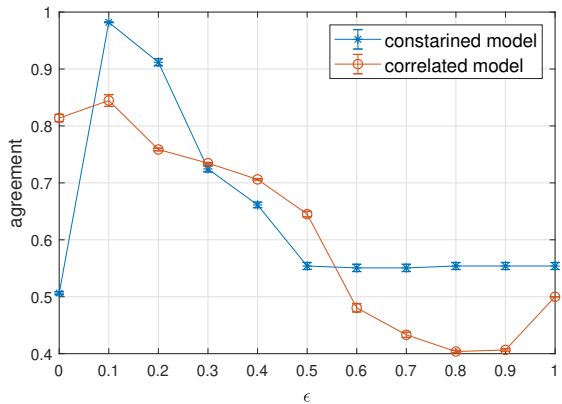}
		
	\end{minipage}
	
	\caption{Detectability transition curves for constrained model and correlated model on heterogeneous networks. $p_{same}$ is 0.9 for the correlated model. The results are averaged over top 20 trials where most node labels satisfy WPP. Error bars represent the standard error of the trials. Refer to Fig. \ref{fig:hete} for synthetic network setup.}
	\label{fig:consVScorr}
\end{figure}

In Fig. \ref{fig:consVScorr}, for both constrained model and correlated model, we examine the agreement score $\frac{1}{N}\sum_i{\delta_{t_i^*(l),\pi(\hat{t}_i(l))}}$ between prediction and ground truth. That is because in this more complex experiment, it is not as straightforward to define a normalized agreement score $Q$ as in Fig. \ref{fig:diff_factors} and Fig. \ref{fig:singleVScorrVSconstr}. \NEW{As stated above, not every trial will converge to the correct point, we therefore select for both models the top 20 trials that satisfy WPP better (without using ground truth information). Specifically, for each trial we count how many pairs of nodes satisfy WPP locally, by calculating $f_{check}$ function over the inferred labels of pairs of nodes.} We observe in Fig. \ref{fig:consVScorr} that for $\epsilon\in[0.1,0.2]$, the agreement score of the constrained model is remarkably higher than the correlated model. The performance advantage benefits from a high fraction of WPP-satisfying results using the constrained model for $\epsilon\in[0.1,0.2]$, as shown in Fig. \ref{fig:ratio}. When this benefit vanishes, for $\epsilon\in [0.3,0.5]$, the constrained model gets similar or worse agreement score than the correlated model. Note that again, the proposed constrained model does not utilize the knowledge that the first 100 nodes have correlated community labels, while the correlated model is supplied with this prior information. The reason of the better performance for $\epsilon\in[0.1,0.2]$ is that, the constrained model manages to fuse information for the first 100 nodes in two layers, meanwhile leaving the remaining 100 nodes intact, while the correlated model tends to unify the entire community structure in the two layers, hence corrupting the remaining 100 nodes. The poorer performance of the constrained model in the noisier $\epsilon\in [0.3,0.5]$ range, we suspect, is due to an optimization in stability caused by many more constraints and factor nodes in the graphical model. On the other hand, the correlated model has a simpler form and is less susceptible to the stability issue.

\NEW{In this section, we have compared our constrained model with a basic correlated model, and results show a higher modeling capability of the constrained model, in presence of heterogeneous community structures. Although the correlated model is simpler, the assumption of a uniform label correlation between two layers does not naturally generate multiplex networks with diverse relations, where only a portion of communities are correlated or consistent. Hence, the correlated model (similar to \cite{Ghasemian2015}) is more appropriate for smoothly evolving temporal networks, and the constrained model we proposed is typically suitable for multiplex networks with different types of relations, where the layers are not necessarily uniformly correlated. In principle, the basic correlated model can be extended so that different nodes can have their own interlayer correlation, and the flexibility of the correlated model can be much greater. However, we expect inference difficulty for such model, given the significantly larger number of free parameters, unless the parameters are properly constrained. Such model design will require nontrivial work and is interesting for future works.}

\NEW{Since the goal of the proposed algorithm is that of fusing consistent communities across layers in general networks and of improving detectability, we are not aware of a directly comparable algorithm that is designed for the exact same goal. Nevertheless, we provide in passing a comparison with a popular multilayer community detection algorithm, Genlouvain \cite{jutla2011generalized}, which maximizes a multilayer modularity function. For the same experiments in this section, when $\epsilon=0.2$, Genlouvain converges correctly only 3 out of 100 trials, while our proposed algorithm has over 40\% success rate. Genlouvain performs similarly to the correlated model in this particular test. The reason is that Genlouvain requires interlayer coupling parameters, which, when not given, and can only be assumed to be uniform. In contrast, our proposed constrained model implicitly infers interlayer coupling through $f_{check}$ factor nodes.}

\subsection{Impact of a known number of communities $q$}\label{effect_q}
For a single layer network, any $q$ that is larger than or equal to the actual value will fit the model well. For example, by setting $q=3$ while performing the BP algorithm in a network generated from SBM with 2 communities, we are allowing each node to choose from 3 distinct community labels. However when the messages have converged, generally most nodes will tend to choose from only 2 of the labels, leaving one barely used. Therefore, the general practice is to opt for a larger $q$, until the free energy of the model stops decreasing \cite{Decelle2011}.

This is in contrast to the constrained multiplex networks. In the experiment of a homogeneous multiplex network, only $q=2$ gives the best performance according to the detectability transition curve. To show this effect, we generate such a 2-layer, 2-community network, with high noise $\epsilon=0.35$. (The noise is so high that when we perform BP algorithm on one of the layers with $q=3$, the community detection is affected and all 3 labels may have a significant presence among nodes, making the decision of $q$ difficult.) Then we run the algorithm with $q$ being 2,3, and 4.
\begin{figure*}[htb]
	
	\begin{minipage}[b]{1.0\linewidth}
		
		\includegraphics[width=16.5cm]{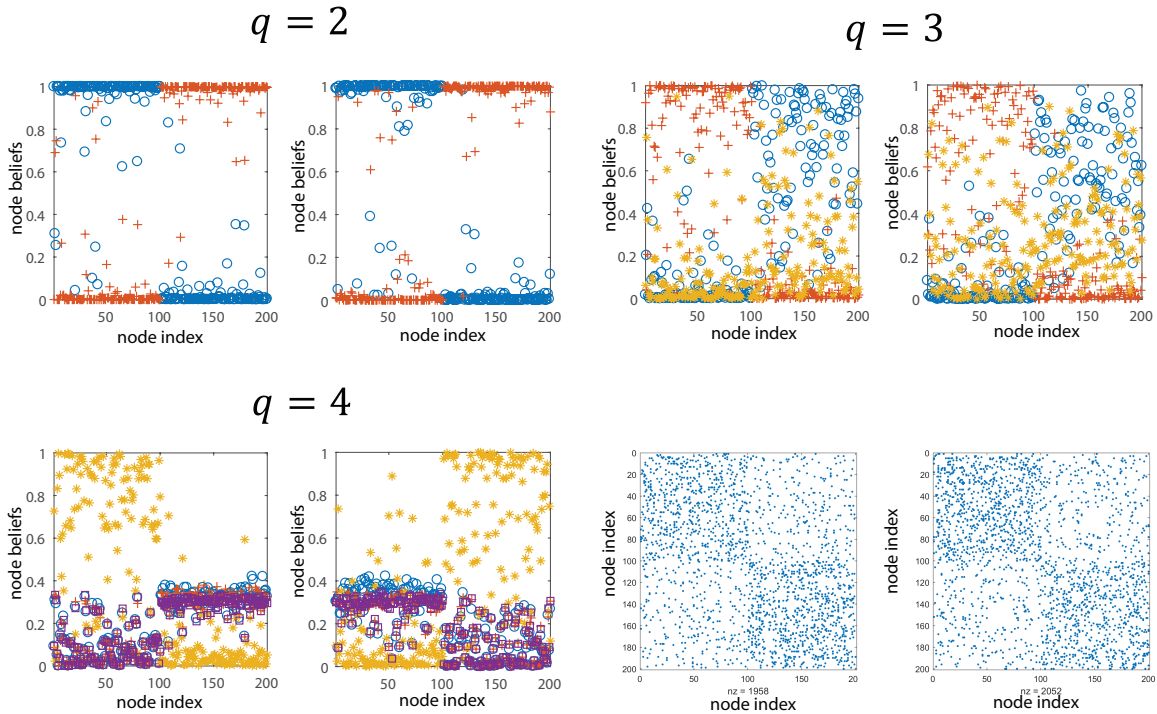}
		
	\end{minipage}
	
	\caption{Probabilities of nodes being assigned certain community labels (node beliefs) at $q=2,3$ and 4. Various community labels are represented by different colors and markers. The lower right image shows the noisy double layer network used in this test.}
	\label{fig:qeff}
\end{figure*}

As shown in Fig. \ref{fig:qeff}, the performance is getting poorer as $q$ increases. Specifically, at $q=2$, most nodes have close-to-one probability of some label, and the selected labels match well among two layers. For $q=3$, the labels still tend to match across layers, but for nodes from 101 to 200, two labels are competing with each other (blue circles and yellow asterisks). For $q=4$, even the labels are not correctly matched. This is because the constraint factors, more specifically WPP, allows the same communities in two layers to be assigned different labels when $q>2$. We therefore cannot combine their information to increase the signal-to-noise ratio. The fact that using the correct $q$ will give a distinctive performance, also enables us to more reliably select $q$.

\subsection{Practical considerations and more layers}

\begin{figure}[tb!]
	
	\begin{minipage}{1.0\linewidth}
		
		\includegraphics[width=8.5cm]{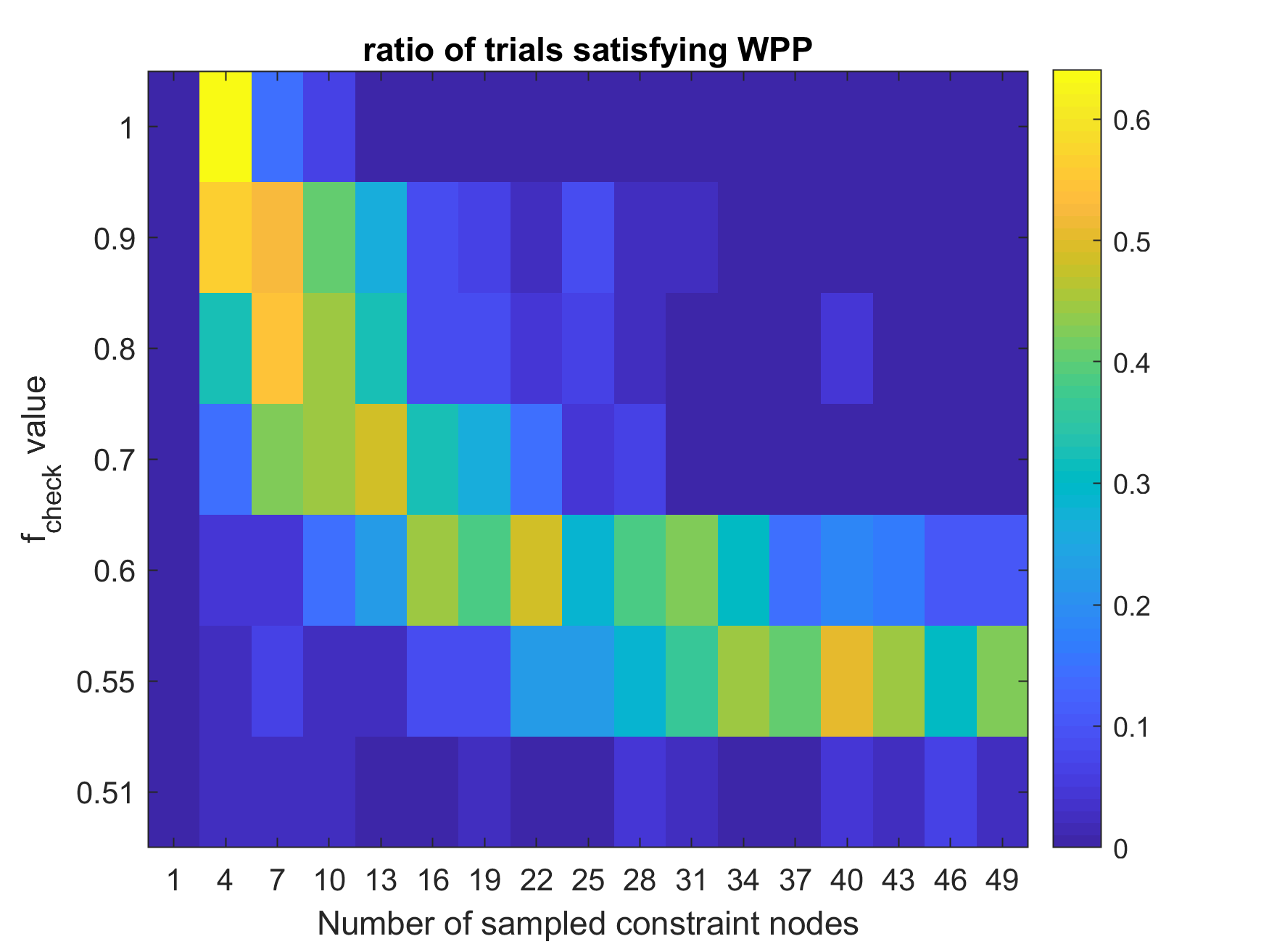}
		
	\end{minipage}
	
	\caption{Ratio of trials that satisfy WPP constraint after convergence, in the parameter space of $f_{check}$ and $N_{sample}$. We identify a band in the parameter space where the constrained model have over half the chance to reach the correct point. The numerical experiment includes 50 trials for each point, and uses the same setup as the heterogeneous two-layer experiment in Section \ref{hete_exp}, when $\epsilon=0.2$.}
	\label{fig:ratio_parascan}
\end{figure}

\NEW{A common challenge in belief propagation algorithm for general graphical model is the presence of a fair number of short loops. Specifically, in our model, the interlayer factor nodes introduce many short loops in our factor graph, both within layer and between layers. These short loops result in a quick convergence to undesirable points, and message update equations become more approximate, due to the influence of $f_{check}$ being overly amplified. To cope with this, we slightly modify message update equations. Specifically, instead of making the product over all incoming messages from neighboring interlayer factor nodes $N_{inter}(i)$, we sample and multiply a fraction $N_{sample}$ of incoming interlayer messages, which also conveniently reduces the computational load. Meanwhile, we can also change the values of $f_{check}$ function from $\{0,1\}$ to, for example, $\{0.2,0.8\}$, to relax the constraint. By applying these modifications, we observe a more reliable and stable convergence to the correct point in our experiments. In Figure. \ref{fig:ratio_parascan}, we find multiple combinations of learning parameters $f_{check}$ value and $N_{sample}$, where the constrained model has over $50\%$ chance to converge to the correct point. These points form a continuous band in the parameter space.}

\NEW{Generally for networks with $N$ node, $L$ layers and $E$ edges per layer, the number of message passing per epoch is $2EL+2(N^2-N)(L^2-L)$, which is dense. For multiplex networks with $L>2$ layers, our original idea needs in total $(L^2-L)/2$ different interlayer factor nodes between pairs of layers, since we do not assume sequential layers. Viewing from the scale of layers, messages between all pairs of layers form high level loops, making it even more difficult to converge correctly. We address this difficulty by adopting the idea of alternating projection. Specifically, in each iteration, we optimize messages in every two layers at a time, while freezing other layers, until all pairs are updated. In this way, we break the high level loops among the layers, and decompose the problem $(L>2)$ into several subproblems $(L=2)$, which are more studied and have better convergence behavior. Another possibility to reduce the complexity is to incorporate this complex structure into a single factor node, an extended constraint function $f_{check}(t_i(1),...,t_i(L),t_j(1),...,t_j(L))$ that covers all the layers at once, instead of just two layers, so the number of interlayer messages will scale linearly to the number of layers. We show an experiment of a 3-layer heterogeneous network to compare these two strategies. The 3-layer network has 90 nodes in each layer, including 5 different communities in total, while a common one exists between layer 1 and 2, and between layer 2 and 3 respectively. We find that only optimizing two layers at a time has a significant advantage in improving the speed and chance of convergence to a correct point (48 correct convergence out of 100 trials). Similar to the experiment in Section \ref{hete_exp}, the correlated model will fail to deal with such heterogeneous structures.}

\section{Conclusions}
\label{sec:conclusion}
We developed a belief propagation algorithm for community detection in general multiplex networks. We considered a case where natural label constraints exist. This case corresponds to a potentially heterogeneous community structure for different layers, a likely scenario for real-world networks. \NEW{As a comparison, we also considered a correlated model where community labels are uniformly correlated across the layers, for homogeneous multiplex networks. Relying on Bayesian inference, our method is theoretically optimal for networks described by our proposed probability model. For the correlated model, combining information from two layers significantly improves detectability due to the additional prior information. More importantly, for the label constrained model, we showed that using just label WPP constraints and limiting the number of communities, we can achieve a similar performance improvement as that of the correlated model, without rather restrictive prior assumptions. Furthermore, the constrained model is able to assign correct labels to heterogeneous commnuity structures, and achieve a much better detection accuracy than the correlated model over some parameter space. This is especially beneficial for detecting sparse and noisy communities in multiplex networks, such as social networks and biological neural networks. Our current constrained model assumes a homogeneous structure within each community. For networks with specific topologies, we can apply modified SBM in our model, such as degree-corrected SBM for social networks \cite{Karrer2011,Newman2015}. Future directions also include improving factor graph design and interlayer message passing efficiency, and applications to real world networks, with the proper numerical efficiencies.}

% or


% \appendices
% \section{Proof}

%\section{}

\ifCLASSOPTIONcompsoc
  \section*{Acknowledgments}
\else
  \section*{Acknowledgment}
\fi

We thank Han Wang for helpful discussions. We would like to acknowledge the support of U.S. Army Research Office: Grant \# W911NF-16-2-0005.

\ifCLASSOPTIONcaptionsoff
  \newpage
\fi



\bibliographystyle{IEEEtran}
% \begin{thebibliography}{40}

\bibliography{huang}% Produces the bibliography via BibTeX.
% \end{thebibliography}


\begin{IEEEbiography}{Yuming Huang}
%\begin{IEEEbiography}[{\includegraphics[width=1in,height=1in,clip,keepaspectratio]{Yuming}}]{Yuming Huang}
received the B.Sc. in Applied Physics from University of Science and Technology Beijing in 2013. He is a PhD. candidate in Physics from North Carolina State University, where he also got his M.Sc. degree en-route. His research include network science, algorithms on networks, computational neural science and machine learning.
\end{IEEEbiography}

 \begin{IEEEbiography}{Ashkan Panahi}
%\begin{IEEEbiography}[{\includegraphics[width=1in,height=1in,clip,keepaspectratio]{Panahi}}]{Ashkan Panahi}
Was a U.S. NRC postdoctoral researcher at NCSU in North Carolina. He received recspectively his MS and Ph.D. in Communication System Engineering and Signal Processing from the EE Department at Chalmers University, Sweden in 2010, and 2015. He received his BS in EE in  2007 from Iran University of Science and Technology, Tehran, Iran. Further, he was also  a postdoctoral fellow  at the Computer Science Department, Chalmers University where he is currently an Assistant Professor. His research includes the development and application of optimization algorithms for various signal processing and machine learning tasks for large amount of data, especially computer vision and image processing.
\end{IEEEbiography}

 \begin{IEEEbiography}{Hamid Krim}
%\begin{IEEEbiography}[{\includegraphics[width=1in,height=1.25in,clip,keepaspectratio]{Hamid}}]{Hamid Krim}
received the B.Sc. and M.Sc. and Ph.D. in ECE. He was
a Member of Technical Staff at AT\&T Bell Labs,
where he has conducted research and development
in the areas of telephony and digital communication systems/subsystems. Following an NSF Postdoctoral Fellowship at Foreign Centers of Excellence,
LSS/University of Orsay, Paris, France, he joined the
Laboratory for Information and Decision Systems, MIT, Cambridge, MA, USA, as a Research Scientist and where he performed/supervised research.
He is currently a Professor of electrical engineering in the Department of
Electrical and Computer Engineering, North Carolina State University,
NC, leading the Vision, Information, and Statistical Signal Theories and
Applications Group. His research interests include statistical signal and image
analysis and mathematical modeling with a keen emphasis on applied problems
in classification and recognition using geometric and topological tools. He has
served on the SP society Editorial Board and on TCs, and is the SP Distinguished
Lecturer for 2015-2016.
\end{IEEEbiography}

 \begin{IEEEbiography}{Liyi Dai}
%\begin{IEEEbiography}[{\includegraphics[width=1in,height=1.25in,clip,keepaspectratio]{Liyi}}]{Liyi Dai}
(S93M93SM13F14) received the B.S. degree from Shandong University, Shandong, China, in 1983, the M.S. degree from the Institute of Systems Science, Academia Sinica, Beijing, China, in 1986 and the Ph.D. degree from Harvard University, Cambridge, MA, USA, in 1993. His research interests include computer vision, machine learning, data analytics, robotics, braincomputer interfaces, control, and operations research. He has authored/coauthored 88 journal and conference publications and is the author of Singular Control Systems (Springer-Verlag,
1989). He received the NSF CAREER Award. He has served as an Associate Editor of the IEEE TRANSACTIONS ON AUTOMATIC CONTROL, an Associated Editor for the Frontiers in Robotics and AI: Sensor Fusion and Machine Perception, and the Co-Chair of the SPIE Independent Component Analyses, Compressive Sampling, Large Data Analyses, Neural Networks.
\end{IEEEbiography}


%\newpage



%\vfill


\newpage

\appendix[Relation of Local Constraints and WPP]
% \section*{Relation of Local Constraints and WPP}
In this section, we mathematically prove our previous claim that the local constraints in Section 2.2.2 are equivalent to the WPP. Take an $L-$layered multiplex  network consisting of a sequence of graphs $G_l=(V,E_l)$ for $l=1,2,\ldots,L$ with the same set of $N$ nodes $V=\{v_1,v_2,\ldots,v_N\}$ and different set of edges $E_l\subseteq V\times V$. According to WPP, the definition of the communities is independent of the layers. Hence, we consider a family $\mathcal{C}=\{C_1,C_2,\ldots,C_Q\}$ of $Q$ subsets $C_q\subset V$ of nodes as the communities. The presence of communities at different layers is represented by a \emph{community structure}  $\mathcal{S}$ , which is a sequence $\mathcal{S}=(S_1,S_2,\ldots, S_L)$, where $S_l\subset\mathcal{C}$ is the subset of clusters being present at the $l^{\mathrm{th}}$ layer. Now, we may define WPP in the following way:
\begin{definition}
A triple $(V,\mathcal{C},\mathcal{S})$ of community structures is said to satisfy the well partitioned property (WPP). If,
\begin{equation}
    \forall l\in[L], C_a,C_b\in S_l;\quad  C_a\cap C_b\neq\emptyset\to C_a=C_b
\end{equation}
\end{definition}
We further make the following definition:
\begin{definition}
A triple $(V,\mathcal{C},\mathcal{S})$ of community structures is said to be "observable" if each community $C_a\in C$ appears in at least one layer, i.e. $C_a\in S_l$ for some $l\in[L]$.
\end{definition}
We take the community assignments $t_i(l)$ as defined in Section 2.2.2. Then, we have the following theorem:
\begin{theorem}\ 

\begin{enumerate}[leftmargin=*]
    \item A set of community assignments $t_i(l)\in [Q]∪ \{\emptyset\}$ fir $i\in[N]$ and $l\in [L]$ corresponds to a community structure $(V,\mathcal{C},\mathcal{S})$ satisfying WPP if and only if:
    \begin{eqnarray}\label{eq:const1}
        &\forall i, j \in [N], l, k \in [L];\nonumber\\
        &(t_i(l) = t_j(k) \neq \emptyset) \to (t_i(k) = t_j(k))
    \end{eqnarray}
    \item The community assignment uniquely identifies observable community structures.
\end{enumerate}
\begin{proof}
For the ``only if`` part in part 1, take $t_i(l) \in [Q]\cup\{\emptyset\}$ as the natural labeling of
some layered complex $(V,\mathcal{C},\mathcal{H})$. For any arbitrary given indices $i,j,l,k$,  if $t_i(l) = t_j(k) \neq \emptyset$, we
have that $t_i(l) = t_j(k) = q$ for some $q \in [Q]$. This means that $C_q \in H_k$ and $v_i \in C_q$.
Hence, from definition we have $t_i(k) = q$, which proves the ``only if`` part.
For the ``if`` part in part 1, take a labeling  satisfying the
condition above. For each $q \in [Q]$ define
\begin{eqnarray}\label{eq:mid}
    & C_q = \{v_i \mid \exists l \in [L]; t_i(l) = q\}\nonumber\\
    & S_l = \left\{C_q \mid \exists i \in [n]; t_i(l) = q\right\}
\end{eqnarray} 
as in Section 2.2.2.
Now, we show that $\mathcal{C} = \{C_q \neq \emptyset \mid q \in [Q]\}$ and $\mathcal{S} = (S_1,S_2,\ldots,S_L)$ satisfy WPP
complex with  $\{t_i(l)\}$ is its natural labeling. Take two communities $C_a,C_b \in H_l$
where $C_a \cap C_b \neq \emptyset$. Then, we can take $m \in C_a \cap C_b$. Moreover, by definition
there exist nodes $i, j$ such that $t_i(l) = a$ and $t_j(l) = b$. Since $m \in C_a$, there exists
a layer $k$ such that $t_m(k) = a = t_i(l)$. From the assumption in \eqref{eq:const1}, we get that $t_m(l) = a$.
On the other hand, $m \in C_b$ implies with a similar approach that $t_m(l) = b$. We
conclude that $a = b$, which shows that $(V,\mathcal{C},\mathcal{H})$ satisfies WPP. Suppose
that $t_i(l) = q$. Then, by definition $i \in C_q$ and $C_q \in H_l$ which shows that $t_i(l)$ corresponds to $(V,\mathcal{C},\mathcal{H})$. This completes the proof of part 1.

For part 2, simply note that the relations in \eqref{eq:mid} hold for any assignment $t_i(l)$
of an observable community structure.
\end{proof}
\end{theorem}
In the above, we show that the WPP is equivalent to the constraint in \eqref{eq:const1}. Now, we show that this set is equivalent to the set of constraints in Section 2.2.2.
\begin{theorem}
The set of constraints in \eqref{eq:const1} is equivalent to the set of constraints in Section 2.2.2 of the paper.
\begin{proof}
Let us first show that the constraints in 2.2.2 imply \eqref{eq:const1}.  Take arbitrary given indices $i,j,l,k$ and suppose that $t_i(l)=t_j(k)=\alpha\neq\emptyset$. Note that from the constraints of 2.2.2 $t_i(k)=\alpha$, since otherwise it contradicts the last line of the constraints in Section 2.2.2. This proves \eqref{eq:const1}. Now, let us prove the converse. Assuming \eqref{eq:const1}, take again arbitrary given indices $i,j,l,l^\prime$ and denote $t_i(l)=\alpha$ and $t_j(l)=\beta$. If $\alpha=\beta$, then the first line of constraints in Section 2.2.2 must hold since otherwise, exactly one of the two labels $t_i(l^\prime),t_j(l^\prime)$, say $t_i(l^\prime)$ equals $\alpha$. Then, since $t_i(l^\prime)=t_j(l)$, we have from \eqref{eq:const1} that $t_j(l^\prime)=t_i(l^\prime)=\alpha$, which is a contradiction. If $\alpha\neq\beta$, then assuming $t_i(l^\prime)=\beta$ leads to $t_i(l^\prime)=t_j(l)$, which according to \eqref{eq:const1} leads to $t_i(l)=t_j(l)$, which is a contradiction. This shows that $t_i(l^\prime)\neq\beta$. Similarly, we get $t_j(l^\prime)\neq\alpha$, which prove the second line of constraints in Section 2.2.2. This completes the proof.
\end{proof}
\end{theorem}
\end{document}